\documentclass{aa}
\begin{document}
%\onecolumn

\thesaurus{04(04.19.1; 03.20.1; 08.16.4; 13.19.5; 10.19.2)}

%%\thesaurus{23(04.19.1; 03.20.1; 08.16.4; 13.19.5; 10.19.2)} supplement

\title{The ATCA/VLA OH 1612 MHz survey. III. Observations of the 
         Northern Galactic Plane.}

\author{ M.N.~Sevenster\inst{1}, 
         H.J.~van Langevelde\inst{2}, 
         R.A.~Moody\inst{1},
         J.M.~Chapman\inst{3}, 
         H.J.~Habing\inst{4}\and
         N.E.B.~Killeen\inst{3} }

\offprints{M.~Sevenster}

\institute{MSSSO/RSAA, Cotter Road, Weston ACT 2611, 
       Australia (msevenst@mso.anu.edu.au)
  \and
   Joint Institute for VLBI in Europe, 
        Postbus 2, 7990 AA Dwingeloo, The Netherlands
  \and
     Australia Telescope National Facility, P.O.Box 76, Epping NSW 1710,
         Australia 
  \and
     Sterrewacht Leiden, P.O. Box 9513, 2300 RA Leiden, The Netherlands }

\date{Received ; accepted }

\authorrunning{M.~Sevenster et al.}
\titlerunning{The ATCA/VLA OH 1612 MHz survey III}

\maketitle

\input psfig

%%%%% \input allinputs
%%%%%%%%%%%%%%%%%%%%%%%%%%%%%%%%%%%%%%%%%%%%%%%%%%%%%%%%%%%%%%%5
\newcount\levelone    \levelone=0
\newcount\leveltwo    \leveltwo=0
\newcount\levelthree  \levelthree=0
\newcount\levelfour   \levelfour=0
\def\chaphead{}                             % needed for appendix
\def\secno{\chaphead\the\levelone}
\def\subno{\chaphead\the\levelone.\the\leveltwo}
\def\subsubno{\chaphead\the\levelone.\the\leveltwo.\the\levelthree}
\def\subsubsubno{\chaphead\the\levelone.\the\leveltwo.\the\levelthree
                           .\the\levelfour}
\def\newsec{\advance\levelone by1 \leveltwo=0 \levelthree=0 \levelfour=0}
\def\newsub{\advance\leveltwo by1 \levelthree=0 \levelfour=0}
\def\newsubsub{\advance\levelthree by1 \levelfour=0}
\def\newsubsubsub{\advance\levelfour by1}
\def\mnsabs#1{{\mnsbs Abstract}{\mnstt #1}}
\def\absnarrower{\advance\leftskip by \abstractindent}
%         \advance\rightskip by \abstractindent}
\def\titlehang{\hangindent\abstractindent \hangafter 0 \relax}
\newdimen\secskipamount  \secskipamount=1pt
\newdimen\subskipamount  \subskipamount=1pt
\newdimen\bottomtol \bottomtol=0.03\vsize
\def\secskip{\par \ifdim\lastskip<\secskipamount \removelastskip \fi
    \vskip 0pt plus \bottomtol \penalty-250
    \vskip 0pt plus -\bottomtol \relax
    \vskip\secskipamount plus3pt minus3pt}
\def\subskip{\par \ifdim\lastskip<\subskipamount \removelastskip \fi
    \vskip 0pt plus 0.5\bottomtol \penalty-150
    \vskip 0pt plus -0.5\bottomtol \relax
    \vskip\subskipamount plus2pt minus2pt}
\def\subsubskip{\par \ifdim\lastskip<\subskipamount \removelastskip \fi
    \vskip 0pt plus 0.5\bottomtol \penalty-150
    \vskip 0pt plus -0.5\bottomtol \relax
    \vskip\subskipamount plus2pt minus2pt \hskip 10pt}
\long\def\aaabstract#1{\centerline{\null}
   \vskip 1.52cm
   {\absnarrower \noindent {\bf Summary.} #1 \par}
   \oneskip \oneskip}
\outer\def\unnumberedsectionbegin #1 #2\par {\secskip \noindent {{\bf  #1}
\dotfill #2}
    \nobreak \vskip 1pt \noindent}
\outer\def\sectionbegin #1 #2\par {\secskip \newsec \noindent {{\bf \secno\  #1}
\dotfill #2}
    \nobreak \vskip 1pt \noindent}
\outer\def\subsectionbegin #1 #2\par {\subskip \newsub {\subno\ {\rm #1} \hfill
#2}
    \nobreak \vskip 1pt \noindent}
\outer\def\subsubsectionbegin #1 #2\par {\subsubskip \newsubsub
    {\subsubno\ {\it #1} \hfill #2}
  \nobreak \vskip 1pt \noindent}

\newcount\eqnumber \eqnumber=0
\def\new{{\rm\chaphead\the\eqnumber}\global\advance\eqnumber by 1}
\def\eqskip{\vskip .5truecm \hskip 5truecm }
\def\eqend{\vskip .5truecm \noindent}

\newcount\fignumber \fignumber=0
\def\nfig{\chaphead\the\fignumber\global\advance\fignumber by 1}
\def\ntab{\chaphead\the\tabnumber\global\advance\tabnumber by 1}

\newcount\fononum \fononum=0
\def\nfn{\global\advance\fononum by 1}
\def\fonono{\the\fononum}

\def\bck{\hskip-0.35em}
\def\wisk#1{\ifmmode{#1}\else{$#1$}\fi}
\def\etal{{et al.$\,$}}
\def\mlod{micro--lensing optical depth}
\def\msyr{\wisk{\,\rm M_\odot\,yr^{-1}}}
\def\cse{circumstellar envelope}
\def\cses{circumstellar envelopes}
\def\twcc{colour--colour diagram}
\def\lf{luminosity function}
\def\lfs{luminosity functions}
\def\df{distribution function}
\def\dfs{distribution functions}
\def\lvd{longitude--velocity diagram}
\def\lvds{longitude--velocity diagrams}
\def\lbd{longitude--latitude diagram}
\def\vlos{\wisk{ V_{\rm los}}}
\def\vrad{\wisk{ v_{\rm rad}}}
\def\vexp{\wisk{ V_{\rm exp}}}
\def\pspeed{\wisk{ \Omega_{\rm p}}}
\def\losa{line--of--sight}
\def\losn{line of sight}
\def\losns{lines of sight}
\def\fv{\wisk{ f_{\rm V}}}
\def\gc{galactic Centre}
\def\mum{\wisk{\mu}m}
\def\lsol{\wisk{\,\rm L_\odot}}
\def\msol{\wisk{\,\rm M_\odot}}
\def\rsol{\wisk{\,\rm R_\odot}}
\def\tsol{\wisk{\,\rm T_\odot}}
\def\vsol{\wisk{\,\rm V_\odot}}
\def\usol{\wisk{\,\rm U_\odot}}

\let\rsun=\rsol
\let\msun=\msol
\let\tsun=\tsol
\let\vsun=\vsol
\let\usun=\usol
\def\vlsr{\wisk{V_{\rm LSR}}}                              % Vlsr

\def\lz{\wisk{\, L_{\rm z}}}
\def\degr{\wisk{^{\circ}}}                                % degrees symbol
\let\deg=\degr
\def\decdeg#1.#2 {\wisk{#1^{\,\rm o}\bck.\,#2}\ }
\def\decmin#1.#2 {\wisk{#1^{\,\prime}\bck.\,#2}\ }
\def\arcmin {\wisk{^{\,\prime}\bck}\ }
\def\decsec#1.#2 {\wisk{#1^{\prime\prime}\hskip-0.42em.\hskip0.10em#2}\ }
\def\arcsec {\wisk{^{\prime\prime}}\ }
\def\kms{\wisk{\,\rm km\,s^{-1}\,}}                    % km s-1

\def\kmsr{\wisk{\,\rm km\,s^{-1}\,kpc^{-1}}}

\def\gt   {$\!$\hbox{\tt >}$\!$}
\def\lt   {$\!$\hbox{\tt <}$\!$}
\def\oversim#1#2{\lower1.5pt\vbox{\baselineskip0pt \lineskip-0.5pt
     \ialign{$\mathsurround0pt #1\hfil##\hfil$\crcr#2\crcr\sim\crcr}}}
\def\gsim{\wisk{\mathrel{\mathpalette\oversim{>}}}} % > over \sim
\def\lsim{\wisk{\mathrel{\mathpalette\oversim{<}}}} % < over \sim
\def\eqref#1{\advance\eqnumber by -#1 \chaphead\the\eqnumber
           \advance\eqnumber by #1 }
\def\?{\eqref{1}}
\def\last{\advance\eqnumber by -1 {\rm\chaphead\the\eqnumber}\advance
     \eqnumber by 1}
\def\eqnam#1{\xdef#1{\chaphead\the\eqnumber}}
\def\appendixbegin#1 #2{\eqnumber=1 \def\chaphead{{#1}}
    \levelone=0\leveltwo=0\levelthree=0\levelfour=0\eqnumber=1\fignumber=1
    \vskip\subskipamount\noindent{\ninepoint\bf Appendix #1\ \ \ #2}
    \vskip\subskipamount\noindent}
\def\noappendixbegin#1 #2{\eqnumber=1 \def\chaphead{{#1} }
    \levelone=0\leveltwo=0\levelthree=0\levelfour=0\eqnumber=1\fignumber=1
    \vskip\subskipamount\noindent{}
    \vskip\subskipamount\noindent}
\def\nfig{\chaphead\the\fignumber\global\advance\fignumber by 1}
\def\anfig{\global\advance\fignumber by 1}
\def\ntab{\chaphead\the\tabnumber\global\advance\tabnumber by 1}
\def\antab{\global\advance\tabnumber by 1}
\def\nfiga#1{\chaphead\the\fignumber{#1}\global\advance\fignumber by 1}
\def\rfig#1{\advance\fignumber by -#1 \chaphead\the\fignumber
            \advance\fignumber by #1}
\def\fignam#1{\xdef#1{\chaphead\the\fignumber}}
\def\tabnam#1{\xdef#1{\chaphead\the\tabnumber}}
\def\aa#1 #2 {, {A\&A,}{ #1, #2} }
\def\aal#1 #2 {, {A\&A,}{ #1, L#2}\ }
\def\aas#1 #2 {, {A\&AS,}{ #1, #2} }
\def\aj#1 #2 {, {AJ, }{#1, #2}\ }
\def\apj#1 #2 {, {ApJ, }{#1, #2}\ }
\def\adass#1 {, {1995, In: {Shaw R., Payne H., Hayes J. (eds.) PASPC 77,
   Astronomical Data Analysis Software and Systems IV, } p. #1}}
\def\apjl#1 #2 {, {ApJ, }{#1, L#2 }\ }
\def\apjs#1 #2 {, {ApJS, }{#1, #2}\ }
\def\araa#1 #2 {, {ARA\&A, }{#1, #2}\ }
\def\mnras#1 #2 {, {MNRAS, }{#1, #2}\ }
\def\bargal#1 { {1996, In: {Buta, R., Crocker, D., Elmegreen, B.~(eds.)
      Barred Galaxies, PASPC 91, San Francisco,} p. #1 }}
\def\thrss#1 { {1999, In: Gibson, B., Axelrod, T., Putman, M.~(eds.)
     The Third Stromlo Symposium: The Galactic Halo,
     PASPC 165, San Francisco, p. #1 }}

\newcount\fignumber \fignumber=1
\newcount\eqnumber \eqnumber=1
\newcount\tabnumber \tabnumber=1

\def\bibitm{\bibitem{}}

\def\sgm{$\sigma$}
\def\SRCF{10}
\def\SRCT{2}
\def\Sct{Sect.~}
\def\Eqt{Eq.~}
\def\Fg{Fig.~}

\newdimen\dfwid \dfwid=17truecm
\newdimen\dfwidd \dfwidd=16.5truecm
\newdimen\dfwdd \dfwdd=14truecm

\begin{abstract}
We present observations of the region between 
$ 5^{\circ} \le \ell \le 45^{\circ}$ and
$ |b| \le 3^{\circ}$, in the OH 1612.231 MHz line, taken from 
1993 to 1995 with NRAO's Very Large Array 
\footnote{The National Radio Astronomy Observatory 
is a facility of the National Science Foundation operated 
under cooperative agreement by Associated Universities, Inc. } (VLA).
These observations are the last part of a larger survey, covering 
$ |\ell| \le 45^{\circ}$ and $ |b| \le 3^{\circ}$ , with the 
Australia Telescope Compact Array (ATCA) and the VLA.
The region was systematically observed on
a 30\arcmin$\times$30\arcmin\ grid in ($\ell,b$) and the
resulting coverage was 92\% , with 965 pointings.
We found 286 OH--masing objects, 161 of which are new detections and
207 have reliable IRAS point--source identifications. 
The outflow velocity was determined for 276 sources.
A total of 766 sources were detected in the combined ATCA/VLA survey, 
of which 29 were detected in two regions of the survey.

{\it The source tables and spectra (Table \SRCT\ and \Fg\SRCF\ can
be downloaded from http://msowww.anu.edu.au/~msevenst/pubs.html}

In this article we analyse the data statistically and give 
identifications with known sources where possible.
The ``efficiency'' of this VLA survey is 75\% of that of the 
ATCA Bulge survey. This efficiency was determined by comparing 
the detections in the region where the two surveys overlap.
The completeness-- and error 
characteristics are similar, though less homogeneous, 
except for the much larger errors in the flux densities.
The relatively large surface number density found in the northern disk,
suggests that we can see the Bar 
extending to higher longitudes on this side of the \gc .
\end{abstract}

\section{Introduction}

We have surveyed a large section of the galactic plane in the OH 1612--MHz 
satellite line. The aim of the survey was to sample the stellar 
dynamics in the plane, cutting through the major components of
the Galaxy. Very strong radiation is emitted in the observed OH line 
by, amongst others, OH/IR stars (for a review see Habing 1996). These
are particularly suitable tracers of the stellar dynamics for a
variety of reasons. Most important is that they are easily observable in 
any region of the Galaxy, their line--of--sight velocity can be
determined very accurately and they represent a large fraction of 
the stellar population. This article is the sequel to previous
articles discussing the Australia 
Telescope Compact Array (ATCA) ``Bulge'' ($|\ell|$$<$10\degr ,
Sevenster \etal1997a, Paper I) and ``Disk'' 
($\ell$$<$$-$10\degr , Sevenster \etal1997b, Paper II) parts of the survey.
Here we discuss the results of the northern plane region, between 
$ 5^{\circ} \le \ell \le 45^{\circ}$ and $ |b| \le 3^{\circ}$.
This region was observed with the VLA in New Mexico.

\begin{figure}
%%\beginfigure1
\fignam\COV
\anfig
{\psfig{figure=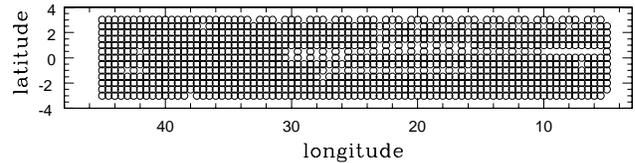,width=9cm}}
\vskip -6.5truecm
%%\caption{{\bf Fig.\nfig }
\caption[]{
The 965 pointings of the VLA survey that were used in the
data reduction; no data were obtained for 88 pointings out of 1053. 
In reality the maps were square (41\arcmin $\times$41\arcmin\ in
$\alpha,\delta$)
and overlapping. A single missing pointing (of which there are
50, cf.~around $\ell$=42\degr )
leaves a hole of $\sim$10\arcmin $\times$10\arcmin\ in the coverage, but 
influences the sensitivity in a larger area.
}
%%\endfigure
\end{figure}

%\beginfigure2
\begin{figure*}
\fignam\RFI
\anfig
{\psfig{figure=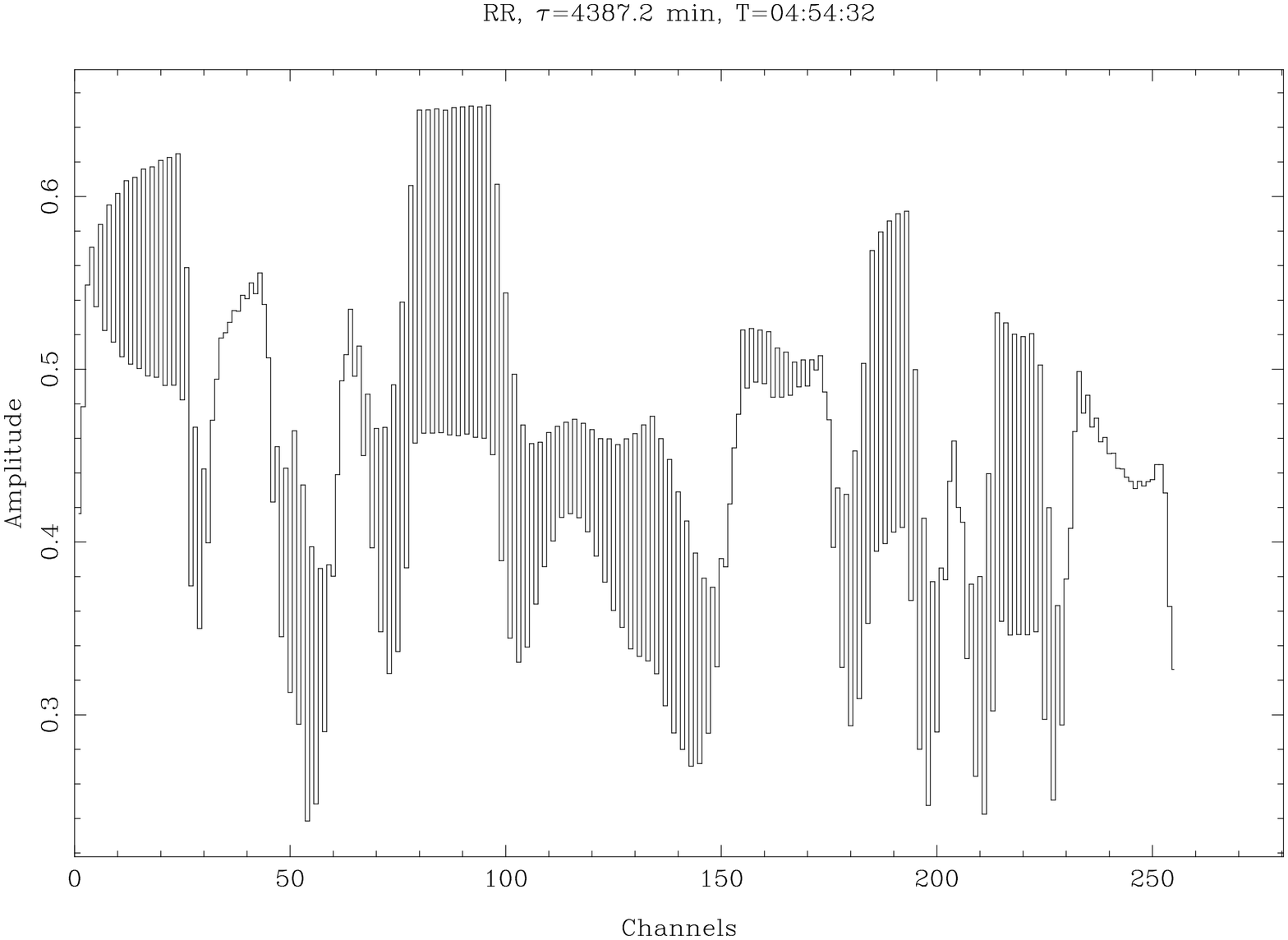,angle=270,height=4.3cm}}
{\psfig{figure=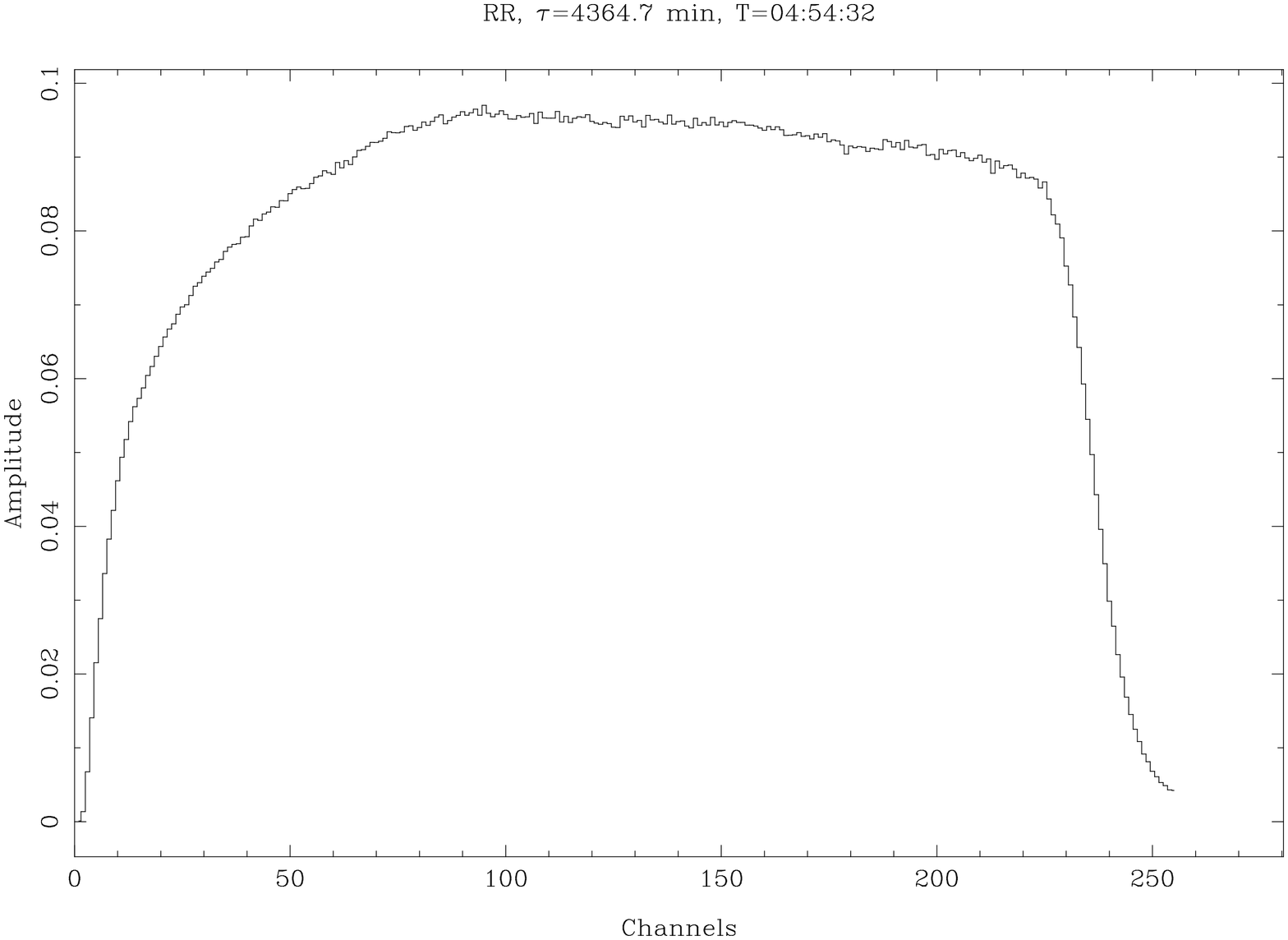,angle=270,height=4.3cm}}
\vskip -8.5truecm
\hskip 6truecm{
{\psfig{figure=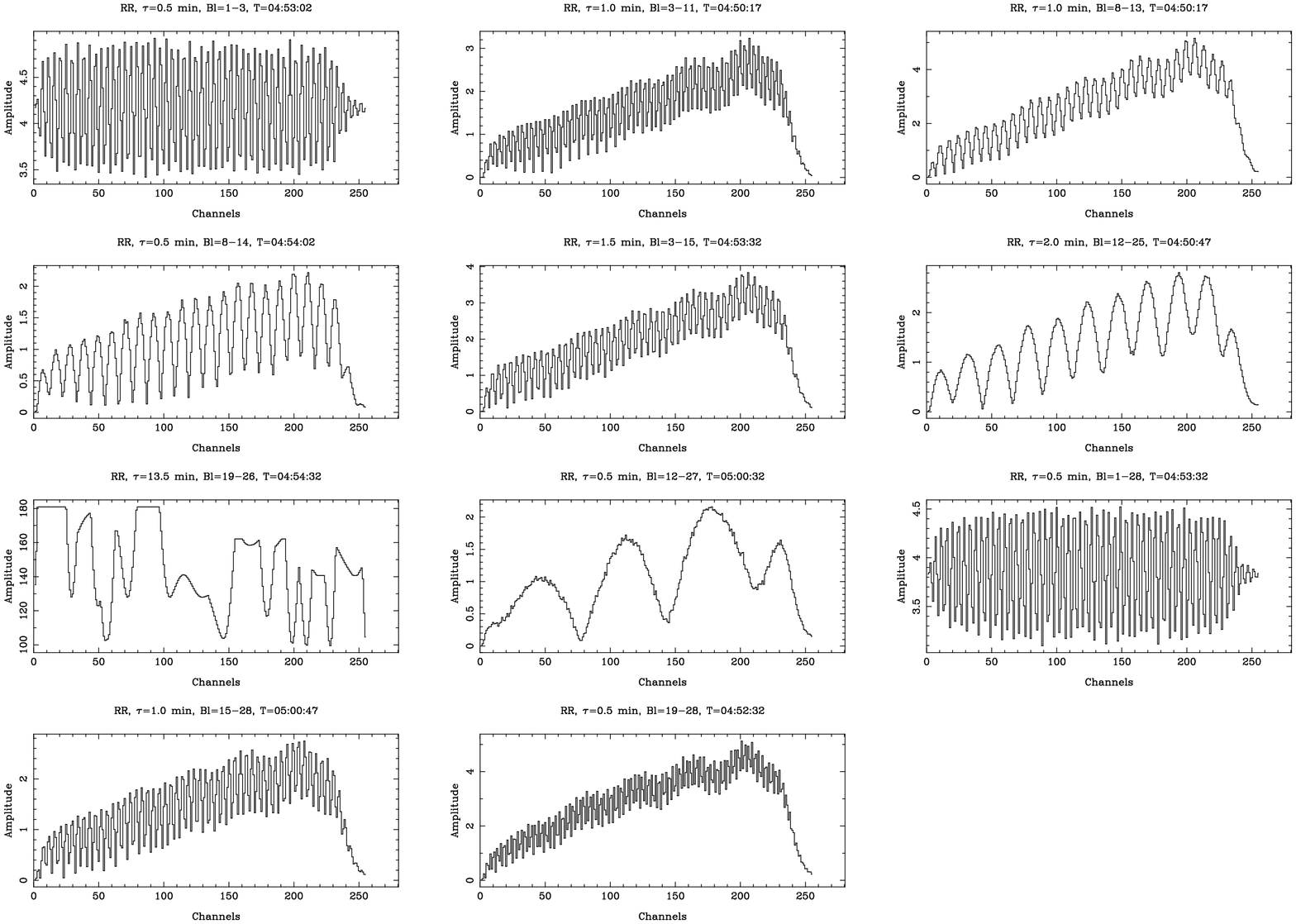,angle=270,width=12cm}}}
%%\caption{{\bf Fig.\nfig a--c. }
\caption[]{{\bf a--c.}
The two larger panels on the left show
the raw spectrum of a primary calibrator (1331+305), taken on 210594,
averaged over all baselines (top) and after flagging with WSRFI (bottom).
The smaller panels on the right show the
single--integration, single--baseline spectra (ie.~visibilities)
that were actually flagged by WSRFI.
Most of these visibilities were flagged as `G' (\Sct 3.1), 
except for the 7th plot, which was a `W'.
}
\end{figure*}

We discuss the observations in \Sct 2 and the data
reduction in \Sct 3. In \Sct 4 we give the results of this survey
and in \Sct 5 a statistical analysis of the data,
that, like Paper II, follows closely the analysis in Paper I.
We summarize in \Sct 6.

\section{Observations}

%%\begintable1
\begin{table}
\tabnam\FIELDS
\antab
\caption[]{All observing runs, with array
configuration, RFI type (\Sct 3.1), maximum baseline length (ranging with
observing direction and time of day) and number of useful fields in the run.}
\tabskip=1em plus 2em minus 0.5em%
\halign to 8cm{
#\hfil & #\hfil & #\hfil & #\hfil & #\hfil & #\hfil & #\hfil \cr
\noalign{\vskip2pt\hrule\vskip2pt\hrule\vskip2pt}
Date & Array & RFI TYPES & UVMAX & N \cr
\noalign{\vskip2pt\hrule\vskip2pt}
200293 & AB & G & 75-110& 73  \hfil\cr
\noalign{\vskip2pt\hrule\vskip2pt}
080693 & BC &  W,G,S$^{206}$ &30-37 & 71 \hfil\cr
\noalign{\vskip2pt\hrule\vskip2pt}
090693 & C &  W,G,S$^{206}$& 16-18& 70  \hfil\cr
\noalign{\vskip2pt\hrule\vskip2pt}
100693 & C & G&17-18 & 70  \hfil\cr
\noalign{\vskip2pt\hrule\vskip2pt}
310893 & C & G,S$^{226}$ & 7-12& 73  \hfil\cr
\noalign{\vskip2pt\hrule\vskip2pt}
020993 & CD & W,G &7-12& 46  \hfil\cr
\noalign{\vskip2pt\hrule\vskip2pt}
090993 & CD & W,G,S$^{226}$&7-12 & 72  \hfil\cr
\noalign{\vskip2pt\hrule\vskip2pt}
%%\noalign{\vskip2pt\hrule\vskip2pt}
210594& AB & W,G,S$^{210}$& 100-120& 38  \hfil\cr
\noalign{\vskip2pt\hrule\vskip2pt}
260594 & AB & W,G & 70-120 & 69  \hfil\cr
\noalign{\vskip2pt\hrule\vskip2pt}
280594 & AB & W,G,S$^{200}_{204}$& 75-120& 73  \hfil\cr
\noalign{\vskip2pt\hrule\vskip2pt}
310594 & AB & W,G,S$_{208}$&  80-120& 62 \hfil\cr
\noalign{\vskip2pt\hrule\vskip2pt}
010694 & AB &  W,G& 75-105& 59  \hfil\cr
\noalign{\vskip2pt\hrule\vskip2pt}
030694 & AB & G& 55& 13  \hfil\cr
\noalign{\vskip2pt\hrule\vskip2pt}
160994 & BC & W,G,S$^{208}_{225}$ & 25-37 & 62  \hfil\cr
\noalign{\vskip2pt\hrule\vskip2pt}
210994 & BC & G,W & 25-37 & 70   \hfil\cr
\noalign{\vskip2pt\hrule\vskip2pt}
220994 & BC &  G,W & 24-37 & 74  \hfil\cr
\noalign{\vskip2pt\hrule\vskip2pt}
210695 & A & W,G& 180-200 & 51 \hfil\cr
\noalign{\vskip2pt\hrule\vskip2pt}
\noalign{\vskip2pt\hrule\vskip2pt}
}
\end{table}

\noindent
The observations were taken with the VLA between
1993 February and 1995 June, in configurations
ranging from A to CD (Table \FIELDS).
The area of this northern part of the survey consisted of
13$\times$81=1053 pointing centres. Due to
system failures and compact antenna configurations in combination
with radio--frequency interference (RFI), 
useful data were obtained for only 965 (\Fg\COV).

The full width at half maximum (FWHM)
of the primary beam (PB) of the VLA antennae at 1612.231 MHz
is 27\arcmin . The data were taken in circular RR polarization, using a total
bandwidth of 3.3 MHz (614 \kms ) and 255 spectral channels
(separation 2.27 \kms ).
Doppler tracking
was used during the observations, transforming the velocities to
the frame of the local standard of rest (LSR), so
the spectral band was centred at 0\kms\ for each pointing.

We used mostly 0137+331, or if necessary 1331+305, as the
primary calibrator and 1822$-$096, 1911$-$201, 1751$-$253 and
1751$+$096 as secondary calibrators.
On each field, an integration time of 10$\times$30 sec was scheduled, 
with in general 300 baselines available for observing.

\section{Data reduction}

Data were reduced with the reduction package 
Miriad (Sault \etal\ 1995); five-- and six--letter acronyms in capitals
throughout this paper will indicate the Miriad programs we used.
To make the data quality uniform across the whole survey area, 
only baselines with lengths between 5 k$\lambda$ and 55 k$\lambda$ were 
used (cf.~Table \FIELDS).
The primary and secondary calibrators were edited first with a
custom--made RFI flagger WSRFI (see \Sct 3.1) and further by hand.
After calibration (MFCAL), the data from all pointings were edited 
only with WSRFI.

The searching was performed largely as described in Paper I, with
MPFND, a custom--made derivative of INVERT, aimed at searching
large amounts of modest--quality data for point sources.
This routine Fourier--transforms spectral channels one by one, retaining
only the position and flux density of 
the brightest pixel for each image. Subsequently, the peaks are
correlated in the spectral direction, looking for statistically 
significant (OH/IR--type)
spectral features. If one has been found, an appropriate point--source model
is subtracted from the visibilities and the same routine is performed
at a lower detection level. All details are given in appendix A of Paper I.

The image size was always 1665$^2$, with square cells of \decsec 1.5 , 
searching all but 10 cells along each border.
Four passes were performed, 
the last one well below 3$\times$ the theoretical noise
(as given by MPFND).

\begin{figure}
\fignam\TISI
\anfig
{\psfig{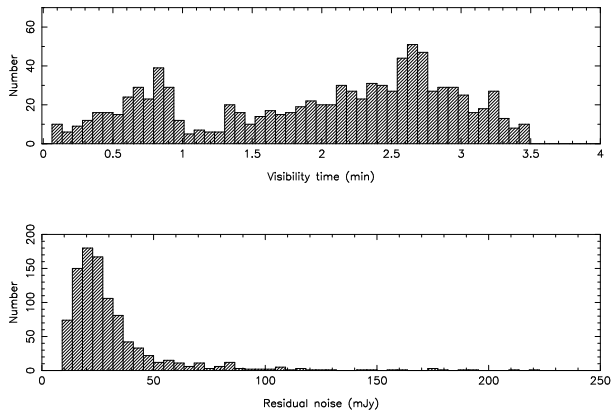}}
%%\caption{{\bf Fig.\nfig a and b.}
\caption[]{{\bf a and b.}
These histograms display the net integration times
and residual noises for all 965 fields 
used in the searching process (see \Sct 2).
For each pointing 5 min integration time was scheduled.
}
\end{figure}

A few changes were made to accommodate specific properties
of VLA data. To save time, with the VLA cubes three times larger
than the ATCA cubes due to higher resolution, an extra 
first pass through the data was performed in the visibility
domain. Spectral peaks were identified in the scalar--averaged
spectra and then only the peak channel was Fourier transformed
to find the source position. 

With the modest spectral resolution of
the VLA correlator, a large fraction of the sources induced
spectral aliassing; the narrowest, mostly brightest, 
sources creating responses in all 255 channels. Obviously, these
reponses may be negative. To account for this, we adapted
the subroutine UVSUB to allow for negative point source models.
Input models for the new UVSUB were ``log'' output files from UVSPEC,
containing the {\it real} 
part of the visibilities, offset to the
source position, for the whole spectral range (215 channels in calibrated data).
Hence, in all passes, once a source position was established,
either directly 
as a real source or after shifting to the real position in the
case of detecting a sidelobe (see Paper I), a full spectral
point--source model was subtracted from the visibility data.
This clearly took out the random noise at the source positions
as well, but the data on the whole were not 
significantly influenced.

The visibility--based point--source subtraction was not as effective for the
VLA data as for the ATCA, due to the non--negligible third dimension
in the antenna--position coordinates (u,v,w) for the VLA.
Sources at higher offsets were occasionally
redetected in later passes. This was corrected for in
the post--searching cross--identification process.

\subsection{Radio--frequency--interference excission}

Three kinds of RFI corrupted the data during most of 
the observing dates. One was the usual broad--band Glonass 
RFI (`G' in Table \FIELDS), often accompanied by a single--channel 
spike (`S', sub--/superscripts indicating channel of spike).
The third was interference from a nearby military base (`W'), 
depending in strength on u--v direction rather than baseline length, which
saturated the correlator to give non--random noise characteristics
(see \Fg\RFI ).

The routine UVLIN, used for the ATCA data (Paper I,II), had no 
positive effect when applied to 
these data with the required high order of polynomial, 
probably again due to the non--negligible third
dimension of the VLA array.
However, with a custom--made visibility--flagging algorithm WSRFI,  written
for use within Miriad, we managed to excise the worst of all three types
of RFI. In \Fg \RFI , we show the spectrum of one of the
primary calibrators before and after running WSRFI on the data.

%Some data examples may be 0335+000 (for W, from 210594) and
%0285+000 (for G, from 210695).

With the combined losses due to antenna downtime, data flagging
to delete interference and retaining only baselines between
5 k$\lambda$ and 55 k$\lambda$, the resulting net `visibility time'
(all visibilities used divided by the number of baselines used, times 30 sec)
is of the order of 1--3 min (\Fg \TISI ).

\section{Results}

{\it The source tables and spectra (Table \SRCT\ and \Fg\SRCF\ can
be downloaded from http://msowww.anu.edu.au/~msevenst/pubs.html}

In Table \SRCT\ all narrow--line OH sources found are listed.
In total there are 286 sources,
125 of which have been identified with known OH--1612--MHz masers.
Ten sources have only one spectral peak.
A reliable (see \Sct 5.5 for definition)
IRAS identification is found for 207 sources.
The median `residual' rms--noise level is 25 mJy (\Fg\TISI , \Sct 5.1.1).

\begin{figure}
\fignam\LBV
\anfig
{\psfig{figure=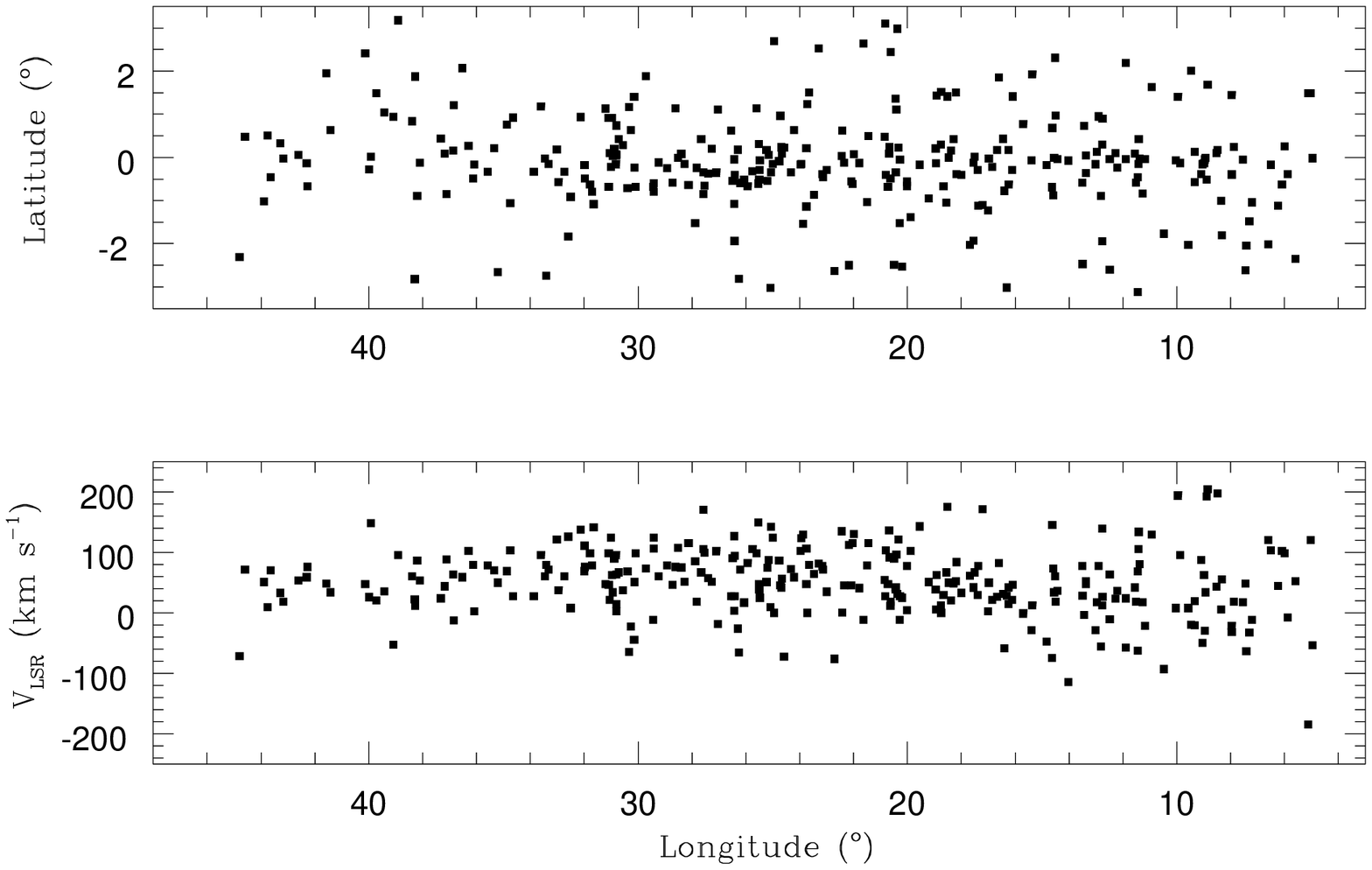,width=8.5cm}}
\vskip -3truecm
%%\caption{{\bf Fig.\nfig a and b.}
\caption[]{{\bf a and b.}
The \lbd\ and \lvd\ for the 286 sources of Table \SRCT .
The central velocities (column 10) are plotted.
}
\end{figure}

\begin{figure}
\fignam\HIS
\anfig
{\psfig{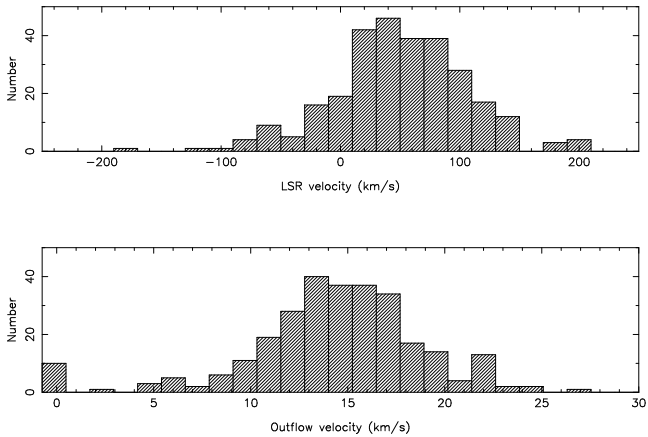}}
%%\caption{{\bf Fig.\nfig a and b.}
\caption[]{{\bf a and b.}
The histograms of the central velocities and outflow velocities
for the 286 sources of Table \SRCT .
}
\end{figure}

For each source the table gives an entry number (column 1),
the OH$\ell-b$ name (column 2), observing date (column 3, see Table \FIELDS),
position in J2000 coordinates (columns 4,5), 
the formal positional error (column 6) -- much smaller than the actual
error (\Sct 5.2) --, the distance from the source
to the pointing centre (column 7), the peak, stellar and
outflow velocities (columns 8 to 11), the peak fluxes (column 12,13),
calibration flag (see \Sct 5.3) and 
error in the peak flux density (column 14; see \Sct 5.3),
the empirical residual noise in the 
field where the source was detected (column 15),
presence of previous OH detection and 
the name of the nearest IRAS point source (column 16)
and the distance to this nearest IRAS point source expressed as a
fraction of the corresponding IRAS error ellipse (column 17).

The spectra for all sources are shown in \Fg \SRCF .
They are displayed with 50 \kms\ on either side of the 
velocity range of each object. 
The spectra were extracted from cleaned (128$^2$ 1\arcsec pixels around
the source position) and restored cubes, summing
over 3$\times$3 pixels around the peak pixel. Only 
spectra \#237 and \#280 were extracted from raw cubes, 
as cleaning was impossible due to the awkward beam shape.
No `continuum fitting' other than the described RFI 
excision (\Sct 3.1) was applied (cf.~spectra \#097,104,253,275,280).

\begin{figure*}
\fignam\LOBE
\anfig
{\psfig{figure=ms10125f6a.ps,angle=270,width=9cm}}
\vskip -6truecm
\hskip 10truecm{
\psfig{figure=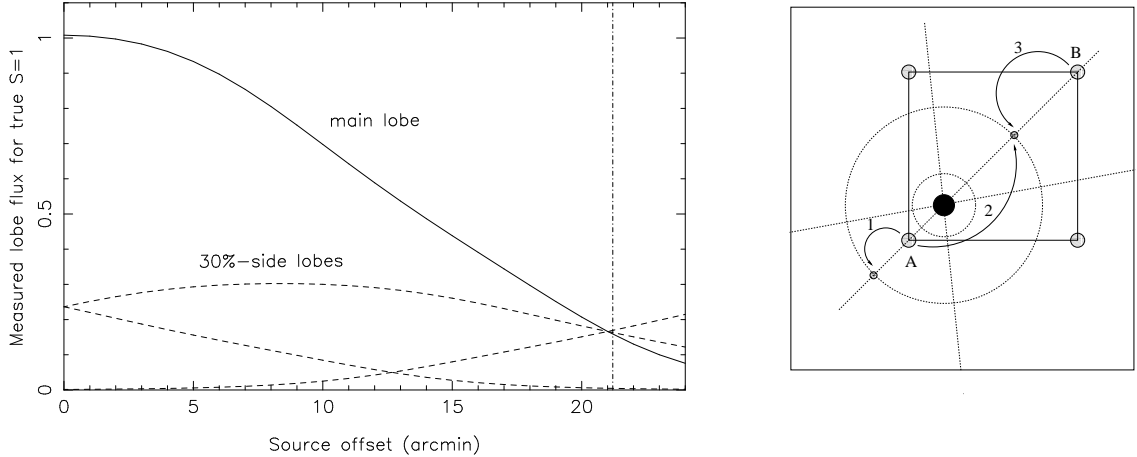,width=5cm}}
\vskip 1truecm
%%\caption{{\bf Figure \nfig }
\caption[]{
A source (solid black circle in right panel)
in the field of pointing centre A will create 
sidelobes in that field, as well as that of B (42\arcmin\ away from A).
The furthest strong (30\%) sidelobe of 
a typical synthesized beam in the VLA survey 
is at $\sim$500\arcsec\ (small hatched circles).
In the left panel, the response to a unity--flux--density source
is given in the main lobe (solid curve) and in this furthest sidelobe, 
which we take to be on a spoke of the beam that runs through A and
B, to calculate the extreme case.
The dashed curves, from top to bottom at offset $<$ 12\arcmin )
give the response in the lobe offset towards A (1), away from A (2)
and the latter seen from B (3), all using 
the known VLA primary--beam attenuation.
The vertical line indicates the largest possible offset of sources
detected in fields that are not on the boundary of the survey region.
Only for sources on the very
edge of a field, the main lobe of the synthesized beam is no longer observed
to be brightest point.
}
\end{figure*}

Dots over peaks in the
spectra give an indication of the velocity range of the
detections; note eg. the sidelobe from spectrum \#223 in 
spectrum \#219. 
Sidelobes are still present in the spectra,
as most confusing sources would be outside the cleaned area.
Negative sidelobes are obvious in eg.~spectra \#127 (from \#132),
\#150 (from \#149) and in an unfortunate way in \#222 (from \#223).
Note the extreme 
%%%positional coincidence of sources \#234 and \#235, and the
velocity coincidence of \#159 and \#172, as well as \#242 and \#244.

In \Fg \LBV\ the \lbd\ and \lvd\ are shown and in \Fg \HIS\ the histograms
of central-- and outflow velocity.

\section{Data analysis}

In this section we analyse the global completeness of the survey
and discuss the statistical accuracy of the quantities given
in Table \SRCT. The discussion and figures follow closely that of the 
corresponding section in Paper I, but is adapted to treat some
VLA--specific details. We will assume errors
are normally distributed, unless stated otherwise.

\subsection{Survey completeness}

\begin{figure*}
\fignam\CMPL
\anfig
{\psfig{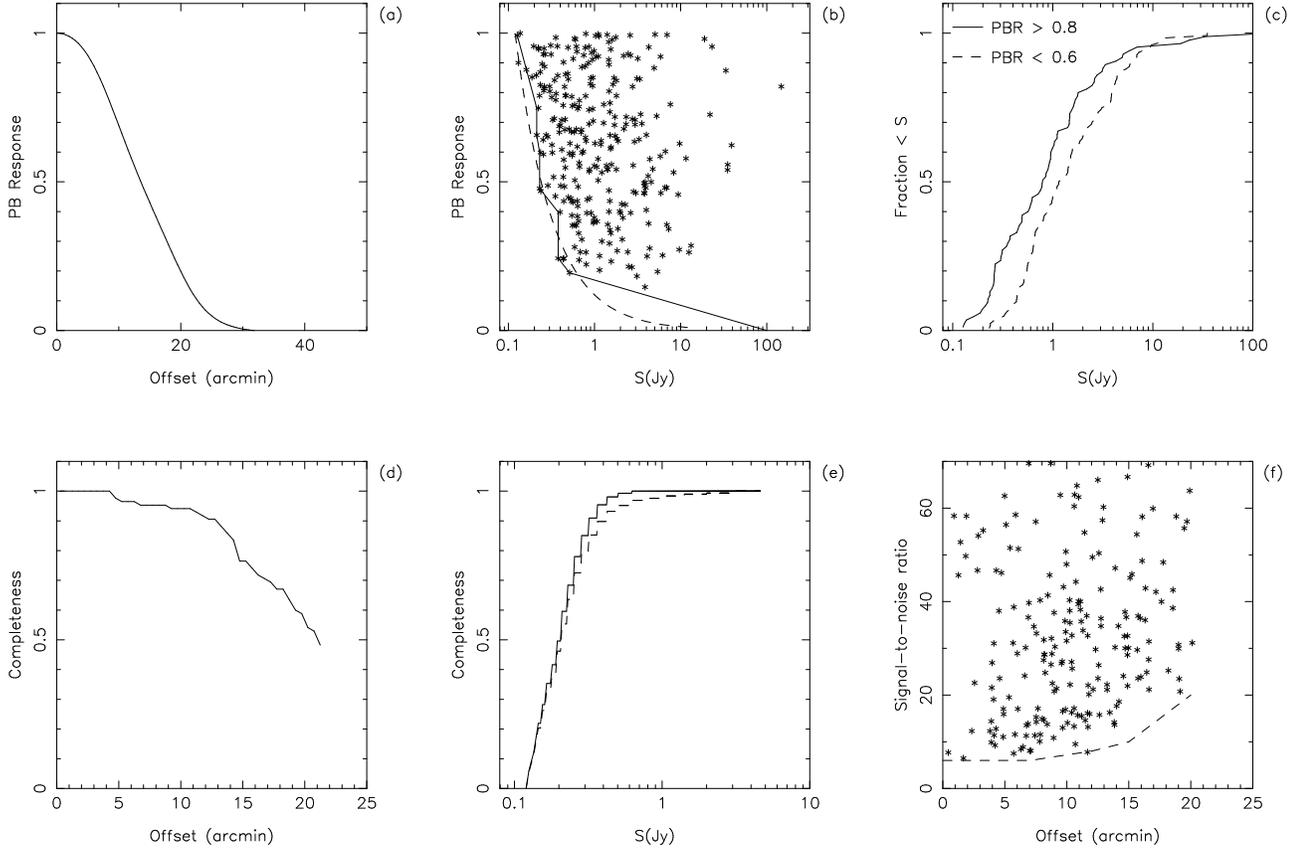}}
%%\caption{{\bf Fig.\nfig a--f. }
\caption[]{{\bf a--f.}
Representation of the completeness of the data (\Sct 5.1.2).
{\bf a} The primary beam response (PBR)
of the VLA antennae at 18 cm, as a function of
radial offset from the pointing centre, taken from
Napier \& Rots (1982).
{\bf b} The PBR calculated for the detected sources
plotted against the highest peak--OH flux densities, corrected for primary beam
attenuation. The solid line indicates the lowest flux densities detected
in PBR bins. The dashed line indicates the expected inner
boundary calculated from the PBR curve in (a) for a limiting
flux density of 120 mJy (chosen to fit observed extremes).
{\bf c} The cumulative flux density distribution for stars with PBR values
$>$ 0.8 (solid line) and $<$ 0.6 (dashed line). The solid line is taken to
be the intrinsic OH flux density distribution for the sources in the survey.
{\bf d} The completeness, relative to the pointing centres,
of the survey as a function of position offsets from the field centres.
{\bf e} The completeness of the sample as a function of flux density.
The offset out to which a source with certain flux density can be
observed is determined from the dashed line in (b). Then we determine
the area covered inside that offset, as a fraction of the total 
area of the survey. The solid curve gives this for the ideal case
of no missing pointings, the dashed curve takes the gaps (\Fg\COV )
into account.
{\bf f} The signal--to--noise ratio for all sources, plotted
against their radial offset from the pointing centre. The dashed
line shows the (observed) lower limit at a certain offset.
}
\end{figure*}

\subsubsection{Noise levels}

The empirical noise levels (defined as the rms in the 
of 215$\times$1645$\times$1645 cubes after the last pass
of source subtraction) for all 965 fields are shown
in \Fg\TISI . The median noise is 25 mJy (mean 32 mJy: the same
as the ATCA Bulge region); 
90\% of the fields have noise levels below 50 mJy. 

\subsubsection{Detection levels}

In this section we treat the global completeness of the sample.
There was no absolute lower detection limit set in the 
searching routines (opposed to the strategy for the ATCA survey).
The half--power beam width (HPBW) of the VLA antennae at 1612 MHz
is \decmin13.5 . Approximately 55\% of the total area of the VLA survey
would be covered within the HPBW (this includes the fact that 8\% of pointings
are missing).
For fields that are not on the boundary of the surveyed region, 
the largest possible offset from the nearest pointing
for any source is \decmin21.2 ,
which corresponds to a primary--beam response (PBR) of 0.16 (\Fg\CMPL a).
In \Fg\LOBE\ we see that this typically guarantees
that the main lobe of a detected source is measured to be stronger than 
any of its sidelobes, even in other fields, provided we cover
the entire survey region with our image sizes 
(square in $\alpha,\delta$ ), which is the case where there are
no gaps (\Fg \COV). Note that the field separation is the
absolute maximum, given the primary--beam response, to still
guarantee proper identification of the main lobes of all sources.
In fields at the perimeter of the
survey region (cf.~\Fg \COV ), sources could be 
found in principle at offsets up to 29\arcmin .
In practice, the largest offset was \decmin21.3 (Table \SRCT ); 
only four sources were found more than 20\arcmin\ away from the 
nearest pointing centre.

In \Fg\CMPL b we plot the PBR of each source against its peak flux density.
The solid line connects stars with the lowest detected OH
flux densities, determined in bins of PBR.
The dashed curve indicates the expected relation between
PBR and flux--density cut--off 
for an absolute detection limit of 120 mJy (cf.~160 mJy, 140 mJy for
the ATCA survey regions). Note that the empirical detection level
is clearly not as constant across the VLA region as across the
ATCA regions.
The global limit is closer to 150 mJy, but there are a few
detections that clearly are outliers (limit 90 mJy). This is due to the gaps 
in the coverage, that increase the area covered by large offsets
from what it would normally be, combined with a larger
spread in the noise levels (see above; \Fg\TISI ).

In \Fg\CMPL c we plot the cumulative flux--density distributions
for all detected sources with PBR $>$ 0.8 (solid)
and, for comparison, with PBR $<$ 0.6 (dashed).
We postulate that the survey is essentially complete for PBR $>$ 0.8,
for flux densities above the sensitivity limit,
and take the solid curve in \Fg\CMPL c as
the intrinsic cumulative flux--density distribution
for the present sample.
We use this distribution to determine the fraction of the 
`flux--density function' seen at a given offset (\Fg\CMPL d), 
with the absolute limit of 120 mJy to determine the cut--off in
flux density at the offset.
In short, the curve in \Fg\CMPL d shows what fraction of
the flux--density distribution we see if we select sources with 
a certain offset from Table \SRCT .\footnote{
Note that the text of Paper I describes this plot incorrectly; 
the area covered by points of that given offset is {\it not} taken
into account. The actual plots in Paper I,II are the same as in the
present paper.}

The solid curve in \Fg\CMPL e shows what fraction of the searched 
area we cover
if we select all sources with a certain flux density from Table \SRCT .
The dashed curve in \Fg\CMPL e shows the same distribution, but this 
time properly corrected for the gaps in the coverage and the corresponding
redistribution of area from small offsets to large offsets.
The dashed curve reaches 99\% completeness for 2 Jy, 
the solid curve for 0.5 Jy. The latter is virtually the same as
the 99\% completeness of the ATCA surveys. The survey is
80\% complete for sources of 320 mJy or 285 mJy, respectively, 
again similar to the ATCA surveys.
This should be interpreted as follows. If we select all sources 
brighter than 2 Jy (after correction for PBR), the distribution
on the sky will be entirely uninfluenced by our observations
and only reflect the real surface--density distribution. If we
select all sources brighter than 0.5 Jy, the distribution will
show holes at the missing pointings, but will not be influenced
by the variable sensitivity within fields. If one selects
even fainter sources, the distribution will be a complicated
function of both intrinsic surface density and survey sensitivity.

Finally, in \Fg\CMPL f, we plot the signal--to--noise ratio for
all detected sources against offset. The limiting signal--to--noise ratio
is 7 (the corresponding ratios for the ATCA surveys are 4 and 6, respectively)
at low offsets. 

Summarizing, the lower sensitivity limit is 120 mJy $\pm$ 30 mJy, with 
the most sensitive detections made at 7$\sigma$.
The survey is 99\% complete for flux densities higher than
0.5 Jy (or 2 Jy, when taking into account holes in the survey area) and
90\% complete in flux density for offsets lower than \decmin 13.2 .
%%%Relative completenesses !

\subsection{Positions}

The positions given in Table \SRCT\ (columns 4,5) are
determined by fitting a parabola over 3$\times$3 1\arcsec$\,$cells
around the peak pixel in the cleaned and restored
map of the peak channel (MAXFIT), made
around the position of the object found by the searching routine.
This ensures that the inaccurate handling by Miriad
of the third coordinate of the visibility domain 
does not influence our final positions.
The errors in Table \SRCT\ (column 6) are the formal errors determined
by fitting a point--source object (IMFIT) to the 
centre of the above--mentioned map.
They are typically negligible with respect to the errors
introduced by the low resolution of some of the observations.

Firstly, the positional accuracy varies dramatically from source
to source, as the maximum baseline length varies from 7 k$\lambda$
(see Table \FIELDS ) to 55 k$\lambda$ , corresponding to a resolution
of 29\arcsec and 4\arcsec , respectively.
With 1\arcsec cells for the imaging of all fields, 
the positional errors for sources observed with 
the lowest resolution would go to infinity (see \Fg 5 in Paper I).
As an example, for a 
maximum baseline length of 25k$\lambda$
the half width of the synthesized beam at half maximum would
be $\sim$8\arcsec , or 5.5 cells, giving an error of $<$2 cells 
or 2\arcsec , following \Fg 5 in Paper I.

Secondly, since we discard baselines below 5 k$\lambda$ (\Sct 3),
for observations with compact array
configurations (Aug,Sep 1993) the synthesized--beam shape
was severely compromised, given the minimal hour--angle coverage
(cf.~\#237,280 with infinite beam size). 
However, our position for \#280 differs only 11\arcsec\ of that
given in the literature (Braz \& Epchtein 1983), 
which is entirely within their errors.

In short, the typical error in the positions 
is of the order of 2\arcsec , in agreement with our findings
in \Sct 5.7.

\subsection{Flux densities}

The flux densities in Table \SRCT\ (columns 12,13)
were determined from the same cubes as described in \Sct 5.2 .
The data were summed over 3$\times$3 
1\arcsec$\,$cells around the maximum cell and divided by the
corresponding sum of the synthesized beam (IMSPEC).
The formal errors (column 14) were determined
by IMFIT, in the peak channel.
Errors marked with $\ast$ (mostly 0.0 mJy) indicate
that the absolute flux--density calibration was not carried out properly 
in the corresponding field.
We applied a fudge factor of 16 to all flux
densities measured in those fields, which introduces
an error of 25\% for those sources (in properly calibrated fields
this factor ranges from 12 to 20).
In all other fields, the error from the absolute 
calibration is of the order of 1\% , except for 
those observed on 260594 where it is 20\% .
From the time--dependent calibration solutions, we expect
errors of 10\% -- 20\% in the flux densities.

\begin{figure*}
\fignam\ALI
\anfig
%\psfig{figure=ms10125f8.ps,width=10truecm}
%%
%%For normal format with caption on left:
\vbox{\psfig{figure=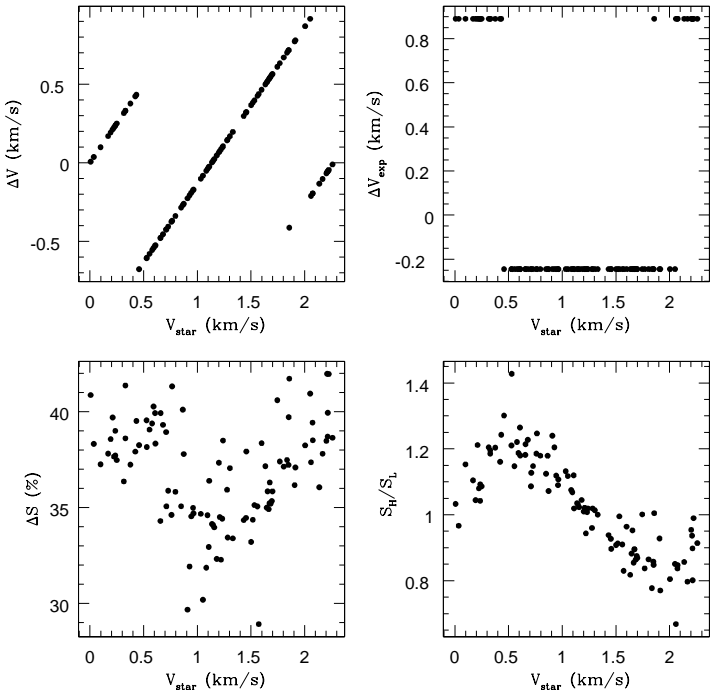,width=10truecm}\vspace{-9.9cm}}
\hfill\parbox[b]{8cm}{\caption[]{{\bf a--d.}
%{\caption[]{{\bf a--d.}
With the channel width of 2.27\kms , the 
intrinsic velocity structure of the maser features,
of order 0.2 \kms\ wide, is undersampled. Hence, aliassing
and measurement errors are introduced, that depend on the
offset of the true spectral peak from the centre
of the channel it is detected in.
This figure gives the difference between `measured' and
`perfect' OH/IR--type spectrum in stellar velocity ({\bf a}), 
outflow velocity ({\bf b}),
flux density ({\bf c}) and peak--flux--density ratio ({\bf d}). The `perfect'
spectrum has an outflow velocity 15 \kms\ and unit 
flux density for both peaks. 
We `observed' it 100 times, adding random noise 
so that the resulting signal--to--noise ratio was between 0.0 and 0.3,
shifting the stellar velocity (abscissae) by random amounts
within one channel width, convolving with 
the spectral--response sinc function and binning to 
the resolution of the observations.
The combined errors are
$\pm$0.8 \kms\ in velocity, mostly $-$0.24 \kms\ in outflow velocity 
(independent of the value of the outflow velocity) and of
the order of $+$35\% in flux density (the measured value is
lower than the true one, mostly due
to the wide bins) with $\pm$5\% due to noise and aliassing.
In the fourth plot we see that the intrinsically symmetric 
spectrum may be distorted by $>$ 20\% .
None of these quantities show
any trend with signal--to--noise ratio.
}}
\end{figure*}

An addional error is introduced by random noise and binning.
In \Fg\ALI c we show that this amounts to
35\% (binning) $\pm$ 5\% (noise). 
The corresponding errors in the peak--flux--density ratio (\Fg\ALI d) is
$\sim$20\% (0.7/0.6 instead of 1/1, \Fg\ALI c).

Summarizing, the relevant errors
are on average 15\% (calibration) plus
5\% (noise) plus for some sources 25\% (fudging).
Multiplying, we find a typical accuracy
for the flux densities in Table \SRCT\ of 20\% ,
which is in agreement our findings in \Sct 5.7 , or 40\% for fudged 
sources.

\subsection{Velocities}

As the VLA has on--line Doppler tracking, the velocity band 
is always centered on 0 \kms\ (LSR), but the bandpass calibration,
in frequency space, changes the velocity range
slightly with position and date. However, all 
detections are within the range that
was covered on all dates ($-$200 \kms , $+$210 \kms ), so
sensitivity and errors are independent of the value of the velocity. 

In \Fg\ALI a we show the difference between measured stellar
velocity, combining the effects of binning, noise and aliassing,
and the true stellar velocity of a perfect spectrum, as a function
of true stellar velocity with respect to the centre of the
channel of detection.
The typical errors are less than 1 \kms , in agreement with 
our finding in \Sct 5.7 .

The same arguments as given in Paper I apply, that for double--peaked
sources the outflow velocity will be systematically 
slightly underestimated (see also \Fg\ALI b) 
due to undersampling of the spectra.
For single--peaked sources there is an 
uncertainty in the {\it stellar} velocity of typically 14 \kms .
The fraction of such single--peak sources in this VLA set is low (3\%)
compared to the ATCA sets in Paper I,II (9\%,18\%).
Due to the variable data quality and remnant RFI, fewer
single--peaked detections were significant according
to our criteria. For instance, a detection in three
separate channels at the same spatial position was more readily 
accepted if it was in two neighbouring channels plus one at a certain
velocity interval than if it was in three neighbouring channels, even
if statistically those two scenarios may have the same probability.
Empirically, the single--peaked configuration is more likely
to turn out not to be a source, but remnant RFI.

\begin{figure}
\fignam\IRASP
\anfig
{\psfig{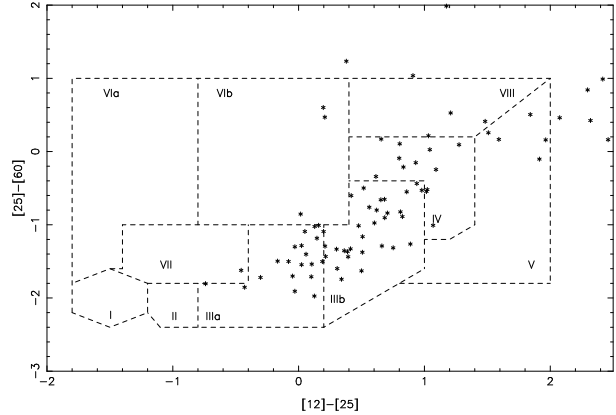}}
%%\caption{{\bf Figure \nfig }
\caption[]{
The IRAS two--colour diagram for sources with an IRAS identification
lying within the IRAS error ellipse (column 17, $N \le$ 1)
with well--determined IRAS 12, 25 and 60 \mum\ flux densities (i.e.
no upper limits).  The colours are defined as
[12]-[25]$\equiv$2.5 $\rm ^{10}$log(S$\rm _{25}$ / S$\rm _{12}$).
}
\end{figure}

\subsection{IRAS identifications}

In column 17, the distance from the OH source to the position
of the IRAS point source (column 16) is given 
in fractional radii of the IRAS error ellipse in the direction of the OH 
position (see Paper I). $N=1$ means the position of the
OH source can be identified
with the position of the IRAS source with 95\% confidence.
The IRAS PS in column 16 is in fact the one that is most likely to
be connected to the OH source, ie. with the smallest value of $N$ .

Of the 286 sources, 207 (72\%) have an IRAS PS counterpart within the IRAS
error ellipse ($N < 1$). This fraction is comparable to those found
in Paper I,II (65\%,75\%), as well as to the 68\% found
by Blommaert \etal(1994).
In \Fg\IRASP\ we show the IRAS two--colour diagram (van der Veen \& Habing 1988)
for sources with an IRAS identification with $N<1$ and reliable IRAS colours.

Note that this two--colour diagram is very similar to that of the
ATCA Bulge OH/IR stars (Paper I), rather than of the ATCA Disk stars,
in terms of the population of region V and the scatter on the 
evolutionary track (regions III). This similarity remains the same
when leaving out the source with longitude below 10\degr . 
It could be due to the orientation of the Bar, that
causes it to extend to higher longitudes in the northern galactic plane.
However, it should be realized that by plotting only those sources that
have well--determined IRAS flux densities in three bands, a selection
effect is introduced. The distribution is influenced by the levels of
background confusion.

\subsection{OH identifications}

We searched the SIMBAD data base for previous OH detections
within 1\arcmin\ from each position in Table \SRCT .
In column 16, as well as in \Fg\SRCF , an `$\ast$' indicates 
a reference to Paper I;
if not in that catalogue, an `x' indicates a reference
to te Lintel \etal(1989); 
if not in that catalogue, a `$+$' indicates another OH reference
and a `$-$' no OH reference at all.
In total, there are 26 `$\ast$' references, which is one less
than determined from the actual Paper I source list (see next section)
since entries in SIMBAD are not always entirely correct.
Note that we did not check for coincidence in velocity.
By this definition, 161 sources in Table \SRCT , or 56\% (cf.~47,55\% for
the ATCA survey regions),
are new detections; see Paper I for a more elaborate discussion.

\subsection{Overlap with the ATCA survey}

The strip of $4^{\circ}.8 < \ell < 10^{\circ}.2$
was observed in both the VLA-- and the ATCA parts of the survey, to
allow direct comparison between the two resulting data sets.
In that region, there are 47 stars detected in the ATCA survey and
36 in the VLA survey, of which 27 stars are actually in common. 
With another two sources in common between the two parts of the ATCA survey
(Paper I,II), the total number of sources in the combined survey is 766.

Taking into account the intrinsic variability of the sources and 
the varying primary--beam responses in systematic observations, even
two surveys with exactly the same flux--density limits would not
yield the same detections unless they were carried out simultaneously
and with the same pointing pattern. 
The redetection 
rate of the VLA is 27/47=57\% and of the ATCA 27/36=75\% .
The difference between those numbers is a direct measure of the relative
``efficiency'' 
( (number of detected sources)/(number of detectable sources within limits)) 
of the two surveys, provided the overlap region is
representative of the whole area. The ATCA
survey is 30\% more ``efficient'' than the VLA survey. 
This is entirely in agreement with the ratio of FWHM coverages : 
the ATCA Bulge survey covers 73\% of the
survey area within a PBR of 0.5, the VLA 55\% , including the missing
pointings.

If we compare this VLA data set to the ATCA Disk data set,
there are 250 and 202 stars, respectively,
in the strips from $\pm$ 10\degr\ to $\pm$45\degr\ in
longitude, even though VLA survey is
less efficient. This could again be interpreted as a sign of
the Bar extending to much higher longitudes on the northern
side of the \gc\ (see \Sct 5.5). 

For the 27 sources in common, the average difference in position
is \decsec1.66 (\decsec0.04 -- \decsec2.91 ) and in velocity 0.68\kms .
The latter value excludes two
sources (\#010=SCHB268, \#021=SCHB286) 
for which one of the two peaks was not detected in the ATCA survey
and source \#020 (SCHB285) which is known to have a very variable 
velocity profile (Sevenster \& Chapman in preparation).

For 8 sources, the observing date (200293) was relatively close to
that of the ATCA, so that a flux--density comparison could be made.
Taking into account the differences in velocity resolution and 
the typical variability of sources, the flux densities are the same
to about 20\% (cf.~\Sct 5.3).

\section{Summary}

We have given the results
of a survey of the region $ 5^{\circ} \le \ell \le 45^{\circ}$ and
$ |b| \le 3^{\circ}$ in the OH 1612.231 MHz maser line.
The survey is 99\% complete for sources brighter than 500 mJy and
90\% complete for positional 
offsets from the pointing centres smaller than 13\arcmin . The absolute
flux density limit is $\sim$120 mJy.
We have found 286 compact OH--maser sources, 161 of which are new detections.
The sources are mainly OH/IR stars, with a few related sources,
like planetary nebulae.
The positions are accurate to 2\arcsec , the velocities
to 1 \kms\ and the flux densities to 20\% .
For 207 sources, an associated IRAS point source is found.
The total number of sources in the combined ATCA/VLA survey 
is 766.

The VLA survey is similar to the ATCA surveys in
terms of global completeness statistics, but considerably less 
homogeneous in quality. The efficiency is approximately 75\% that of
the ATCA Bulge survey. The biggest differences are the considerably 
worse accuracy in the flux densities and the very low fraction of
single--peaked sources.
Signatures of a higher fraction of bar--like stars in this part of
the survey than in the ATCA Disk survey are found 
in the surface number density of detected sources and arguably
in the IRAS two--colour diagram.

%%\section*{Acknowledgments} 
\begin{acknowledgements}

This research has made use of the
Simbad database, operated at CDS, Strasbourg, France.
\end{acknowledgements}

%%\section*{References}
%%\beginrefs

{}

%\vskip 2truecm
%\vfill
\eject
%\onecolumn

\renewcommand\labelitemi{\normalfont\bfseries }

{\bf Table \SRCT} {Compact OH--maser sources in the northern 
galactic Disk region 
(obtain from http://msowww.anu.edu.au/~msevenst/pubs.html)}

The columns of Table \SRCT\ contain the following information :

\begin{itemize}{}
\item{1} Sequence number (coincident with spectra in \Fg\SRCF )
\item{2} Name in the OH$\ell-b$ convention
\item{3} Date of observation. Date `000000' means data were taken 
during more than one observing run
(see Table \FIELDS\ for parameters)

\item{4} Right ascension of the brightest peak
for epoch J2000 (typical error 2\arcsec , \Sct 5.2)

\item{5} Declination of the brightest peak
for epoch J2000 (typical error 2\arcsec , \Sct 5.2)

\item{6} Formal measurement error in position
from IMFIT (determined in the channel of the brightest
peak) in arcsec (\Sct 5.2)

\item{7} Radial offset of the source from pointing centre in arcmin

\item{8} Line--of--sight velocity with respect to the LSR of the
blue--shifted (L) peak.
For single--peaked
spectra the velocity of the peak is always given as blue--shifted
for reasons of tabulation (typical error 1 \kms , \Sct 5.4).

\item{9} Same for the red--shifted (H) peak (typical error 1 \kms, \Sct 5.4).

\item{10} Stellar velocity (typical error 1 \kms / 14 \kms , \Sct 5.4).

\hskip 2.truecm{$ v_{\rm c} =0.5\cdot \left(v_{\rm H}+v_{\rm L} \right) $}\hfill

\item{11} Outflow velocity; zero for single--peaked sources 
  (typical error 1 \kms , \Sct 5.4).

\hskip 2.truecm{\vexp\ $=0.5\cdot \left( v_{\rm H}-v_{\rm L} \right) $ }\hfill

\item{12} Flux density in image domain at peak pixel of cleaned, restored
image (IMSPEC),
corrected for primary--beam attenuation, but not for any `continuum'
of the blue--shifted (L) peak (typical error 20\% , \Sct 5.3).

\item{13} Same for the red--shifted (H) peak (typical error 20\% , \Sct 5.3).

\item{14} Formal measurement error in flux density, determined in the channel
of the brightest peak (IMFIT). An asterisk
indicates that the corresponding field was not calibrated properly (\Sct 5.3)

\item{15} Empirical noise in `empty' cube for the present
field (\Sct 5.1.1, channel width 2.27 km/s)

\item{16} A `$+$',`$\ast$' or `x' for a, or `$-$' for no,
OH identification within 1\arcmin\ in
the Simbad database (\Sct 5.6), and the IRAS PSC position with highest 
identification probability (ie.~smallest $N$, see column 17, \Sct 5.5)

\item{17} Ratio between the size of the error ellipse of, and
and the distance to, the IRAS PS of column 16,
in the direction of the OH position (\Sct 5.5)

\end{itemize}{}

\end{document}